# Investigation of Angle dependent SdH oscillations in Topological Insulator Bismuth

N. K. Karn[1,2,*], Yogesh Kumar[1,2,3], Geet Awana[4,5] and V.P.S. Awana[1,2]

[1]CSIR-National Physical Laboratory, Dr. K. S. Krishnan Marg, New Delhi-110012, India
[2]Academy of Scientific and Innovative Research (AcSIR), Ghaziabad 201002, India
[3]School of Science, RMIT University, Melbourne, VIC 3001, Australia
[4]Visitor School of Physical Sciences, Jawaharlal Nehru University, New Delhi 110067, India
[5]Formerly at Department of Physics, Loughborough University, Loughborough LE11 3TU, UK

**Abstract**

The current article investigated the band structure in the presence and absence of spin-orbit coupling (SOC), examined the Z2 invariants, and investigated the detailed angle-dependent magneto-transport of up to 10 T (Tesla) and down to 2 K for the Bismuth crystal. The out-of-plane field-dependent magnetoresistance (MR) is positive and is huge to the order of $\sim 10^4$% at 2 K and 10 T. On the other hand, the longitudinal (in-plane) field-dependent MR is relatively small and is negative. The thermal activation energy is also estimated by using the Boltzmann formula from resistivity vs temperature measurement under applied transverse magnetic fields. The topological nature of Bi is confirmed by Z2 invariant calculation using Density functional theory. PBESol bands show trivial but Hybrid functional (HSE) bands show non-trivial topology being present in Bismuth. This article comprehensively studies the dependence of MR oscillations upon the angle between the applied field and the current. The observed oscillations fade away as the angle is increased. This article is an extension of our previous work on Bismuth [1], in which we conducted a comprehensive analysis of its structural and micro-structural properties along with its transport behavior in an applied transverse magnetic field.

**Keywords:** Magnetoresistance, SdH oscillations, Single crystal, Berry's phase, LL index, Z2 invariant

**\*Corresponding Author**
Mr. N. K. Karn
National Physical Laboratory (CSIR), Dr. K.S. Krishnan Marg, 110060, India
E-mail: nkk15ms097@gmail.com
Ph. +91-11-45609357, Fax-+91-11-45609310

**Introduction**

Shubnikov-de Haas (SdH) oscillations are among the most intriguing phenomena in magneto-transport measurements, named after Wander Johannes de Haas and Lev Shubnikov. These oscillations are observed in the conductivity of a material, mostly occurring at low temperatures and in the presence of a high magnetic field [2-4]. It also reveals the quantum mechanical behavior of the material, most commonly researched to evaluate the π berry phase in the topological quantum materials [5, 6]. SdH oscillations in conductivity can be used to examine various intrinsic parameters of systems, such as the majority and minority carriers, the effective mass of holes/electrons, and the charge carrier population [2-9]. The main reason for such oscillations in the conductivity lies in their charge carriers, such as electrons and holes, which start cyclotron motion in the presence of a high magnetic field and low temperature. It is also observed in SdH oscillations that the free charge carriers in the conduction band of materials are the main reason for such kind of oscillations [10-12]. The oscillation period varies significantly on the applied field strength, whereas its amplitude varies on both temperature and magnetic field. The magnetic field strength dependence can be interpreted as energy spectra made of confined energy Landau levels. These Landau levels are not continuous; however, these are discrete and separated by a constant amount of cyclotron frequency [13]. Theoretically, these Landau levels are split by Zeeman energy, which is also proposed in the SdH oscillations. As per the literature, the Zeeman energies, cyclotron frequency, and the number of electron states in each Landau level are found to be linearly dependent on the magnetic field [14,15]. Thereby, with an increase in magnetic field, the Landau spin-split levels transfer to higher energy levels. Thus, periodic oscillations are generated at high magnetic fields in the transport conductivity of materials due to oscillations of free charge carriers [12,16]. These SdH oscillations are quite noble and only observable in selective materials [2-16]. Particularly, in the context of Bismuth, these oscillations and magnetoresistance are thoroughly reviewed in [17].

With the discovery of topological quantum materials, which could host the SdH oscillations, the interest of researchers has been regained in these oscillations [18-20]. Topological insulators are quantum materials that offer both insulating bulk and metallic surface characters in a single system [21, 22]. These particular noble characteristics of topological materials have capabilities of showing tremendous advantages in various applications such as spintronics, optoelectronics, thermoelectricity, magnetic sensors and quantum computing [23-27]. There are a few experimental techniques through which the topological character of a material can be directly verified. Angle-resolved photoemission spectroscopy (ARPES) is one such technique [28, 29]; however, its complex measurements and modelling make it quite a vigorous approach to probing topological materials. On the other hand, the magneto-transport measurements provide an indirect method to characterize topological materials. In literature, the presence of topology in Bismuth has been explored, both by showing non-trivial topology via confinement effects [30] and higher-order topology by theoretical calculations [31]. The topology in Bismuth has also been classified based on topological defects [32]. Bismuth has so much to offer in condensed matter physics and possible applications that the same yet continues to fascinate physicists and scientists [30-33]. Interestingly, both theoretically and experimentally, contrasting results have also been found on the topology of Bismuth [34-36]. Our theoretical calculations support the recent resolution being presented by Aguilera et. al. [37].

Therefore, in the present study, we have performed both magneto-transport measurements and theoretical calculations. The Berry's phase of Bismuth and its dependence on the angle between the applied magnetic field and the current is explored. The Z2 invariant is calculated using DFT based method to classify the topology being present in Bismuth. This study provides a complete overview of SdH oscillations and their dependence on magnetic field angle, which can be used further to develop spintronics devices for quantum computing and magnetic sensors.

**Methods**

The computational simulations and experimental measurements of Bismuth single crystal are performed in this article. The Density functional theory-based analysis is calculated by using Quantum Espresso [38, 39] considering the Perdew-Burke-Ehrnzoff (PBESol) type pseudo-potentials to incorporate corrections for exchange-correlation potentials. The bulk electronic band structure and density of states (DOS) are calculated with and without considering the spin-orbit coupling (SOC) effects. For SOC band structure, full relativistic corrected pseudopotentials are used from the PSEUDODOJO library and the calculation parameters are the following: total energy convergence threshold is $5.88 \times 10^{-5}$ Ry/atom, kinetic energy cutoff for charge density and potential is 264 Ry, the energy cut-off for plane wave function is 66 Ry. The self-consistent convergence tolerance is taken as $4 \times 10^{-10}$ Ry. The first Brillouin zone (FBZ) is sampled on an $8 \times 8 \times 8$ mesh generated by the method given by Monkhrost-Pack [40]. For further analysis, the 8 Bloch wavefunctions are wannierized out of 18 obtained from DFT calculation in the WANNIER90 software [41]. The s and p orbital projection are used with disentanglement convergence limit $10^{-8}$. For all calculations, the rhombohedral phase of Bismuth is considered and the lattice parameters are taken from the Rietveld refinement of the grown Bismuth single crystal [1] and then optimised. An effective tight-binding (TB) model for Bi crystal are obtained using Wannier functions. This effective TB model is further processed and applied in Wannier Tools [42] to calculate the Z2 invariant using the Wannier Charge Centre (WCC) evolution in FBZ which are sampled on much denser grid 41×41. The Z2 invariant is calculated by setting surface card in (111) direction and 10 slabs are considered for the calculation.

The single crystals of Bi are grown by the self-flux method by vacuum encapsulating at $10^{-5}$ Torr palletized Bi subjecting to the heat treatment followed in ref [1]. The as-grown Bi crystal is characterized through Single-crystal XRD and Raman spectroscopy is also used to investigate the Raman active vibrational modes. The detailed characterization of as-grown samples can be found in ref [1]. Also, the Quantum Design physical property measurement system (PPMS) is used to investigate field-dependent RT (resistivity vs temperature) and RH (resistivity vs magnetic field) at various magnetic fields and temperatures, respectively. All the details of these techniques are mentioned in the previous article. Here, in this article, the transport properties of Bismuth are further investigated using the PPMS, where the field-dependent RT and RH at different angles are analyzed in order to probe the SdH oscillations behavior with the angle and field. The sample is mounted on rotatory PPMS puck with linear four probe contact. Here, specifically, the addition of magnetic field angle has provided more insights into the phenomenon of SdH oscillations in Bismuth.

**Results and discussion**

The first-principle-based calculations are performed to decipher the electronic and topological character of Bi. In quantum espresso, the Kohn-Sam equation is solved iteratively in a self-consistent way until the solution converges. The electronic properties including bulk-electronic band structure and DOS are also calculated. The self-consistent calculation produces converged electron density functional and Bloch wavefunctions. The calculated Fermi energy is found to be $E_F$ = 10.06 eV. Fig. 1(a) depicts the computed DOS for both the without SOC and with SOC scenarios. It is observed that the calculated DOS is non-zero at the Fermi level, which indicates the metallic behavior of Bismuth crystal in the rhombohedral phase. By considering the SOC effect, it is observed that the DOS peaks get dispersed, which

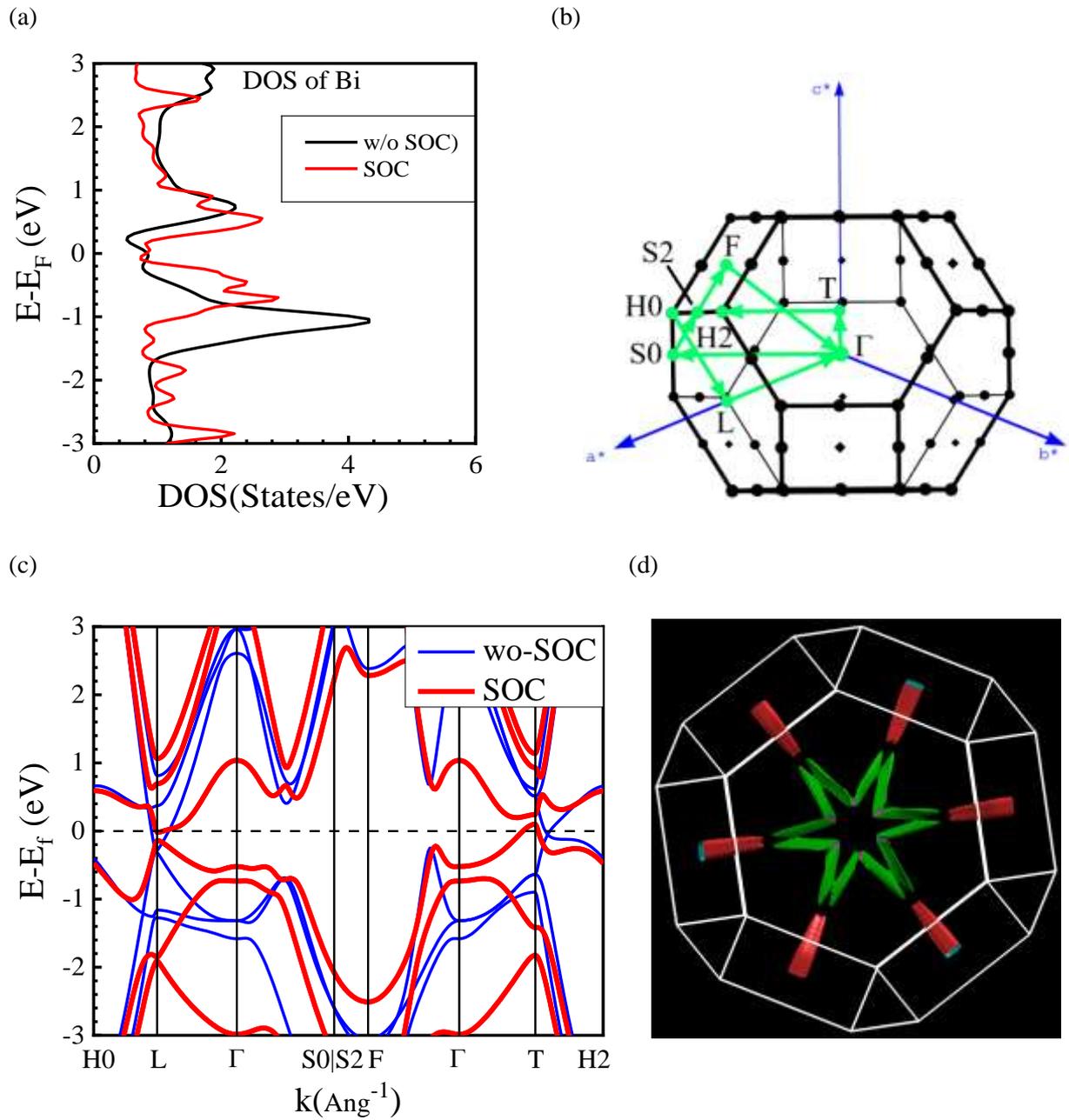

*Fig1. (a) Shows the calculated density of states (DOS) for rhombohedral Bi considering with and without SOC effects (b) Depicts the FBZ with high symmetry point path for band structure calculation guided by green arrow (c) Shows the bulk electronic band structure of rhombohedral Bismuth in reciprocal space k (d) Depicts the Fermi surface corresponding to the two bands crossing the Fermi Level.*

indicates that the band structure of Bismuth is strongly affected by the presence of SOC. To calculate the bulk band structure, Seek k-path is used to find the optimal path based on the previous report by Hinuma et al., [43]. The green arrows in Fig. 1(b) show the path along the high symmetry points in the FBZ of the rhombohedral lattice of Bismuth. As illustrated in Fig. 1(b), the k-space path is H0-L-Γ-S0|S2-F-Γ-T-H2. The calculated bulk band structure along the optimized path is shown in Fig. 1(c) for the without (w/o) SOC and with SOC parameters by blue and red curves, respectively. All bands are plotted with respect to the shifted Fermi level, i.e., $E_F=0$. The band structure calculation also validates the strong SOC, which is expected in Bismuth due to its high atomic number. In the case of w/o SOC, two bands cross the Fermi level and these two bands are degenerate at a point in the path H2-T. When the effect of SOC is included, the band degeneracy is completely lifted. Fig. 1(d) shows the Fermi surface corresponding to the bands crossing the Fermi level. The green surface shows the hole pocket whereas the red surface shows the electron pocket. For the benchmarking and reliability of DFT calculation [44,45] different pseudopotentials are used and corresponding calculated bands are shown in the supplement.

The topological character of materials is elucidated by invariants associated to the symmetry preserved by the system, viz. time reversal symmetry (TRS) broken systems (phase) are characterized by Chern number [46], systems with mirror symmetry are characterized by Mirror Chern number [47] and TRS protected phase has Z2 invariant [48]. The Chern number of the $n^{th}$ band can be computed by integrating the Berry curvature over the first Brillouin zone. For Z2 invariant, Berry curvature integral is over half BZ due to TRS. All these invariant quantities are fundamentally associated to the Berry phase of the system [46]. Since Bi respects TRS, suitable invariant is Z2 for its topological characterization. There are few methods to calculate Z2 invariant: (i) Calculating the band parity/Pfaffians at time reversal invariant momenta (TRIM) points [48] (ii) Fukui-Hatsugai method [49] (iii) Counting the exchange of Wannier charge centers (WCC) when evolved in reciprocal space (FBZ) form loops known as Wilson loops [50,51]. Here, we follow Wilson loops method. From DFT calculated Bloch states are wannierised using the s and p orbital projections, and implemented in WannierTools for the calculation of the Z2 invariant [42]. The SOC bands near/crossing the Fermi level are completely gapped out, which allows us to calculate the Z2 invariant of the same. Based on the windings of WCC in Wilson loops, Z2 invariant is calculated. If there is an odd number of exchanges, a non-trivial (Z2=1) is assigned and for an even number of exchanges, a trivial (Z2=0) value is assigned. Fig. 2 shows the evolution of WCC in six different planes and corresponding Z2 values. Four indices ($\upsilon_0$; $\upsilon_1$ $\upsilon_2$ $\upsilon_3$) are used to characterize the topological insulators given by the formula

$$\nu_0 = \sum of\ Z2 \in k_i\ plane\ mod\ 2$$

$$\nu_i = \sum of\ Z2 \in (k_i = 0.5)\ plane$$

where i=x,y,z. The first index indicates the strong topology whereas the rest indicates the weak topology present in the system. Using the Wilson loop method implemented in WannierTools, the Z2 invariant index for the Bi turns out to be (0,000), which shows the trivial states present in the Bismuth. However, following the above procedure, the Hybrid functional (Heyd-Scuseria-Ernzerhof HSE) band structure produce non-trivial Z2 invariants. The different pseudopotential, results in different band structure, thereby different Z2 invariant is obtained [see supplement]. According to previous reports, both the experimentally observed and theoretically calculated Z2 values are not compatible with each other [34-36]. For example, some of the experimental results (ARPES measurement) [34-35] show non-trivial topological surface states, whereas the theoretical value of Z2 turns out to be trivial [36]. In a recent work [37], this ambiguity is shown to be resolved by including the GW correction in the theoretical calculation. The contradiction is resolved by showing the 'crosstalk' effect, where the surface states with long decay length hybridize in such a way that a band gap is created, which should be a degeneracy otherwise. Here in this article, we also show the presence of non-trivial topology in Bismuth crystal, which is in accordance with Aguilera *et. al.,* [37]. Our theoretical calculation and previous works indicate the existence of topological surface states in Bismuth. For further investigation, we have performed angle-dependent MR measurements on the single crystal of synthesized Bi.

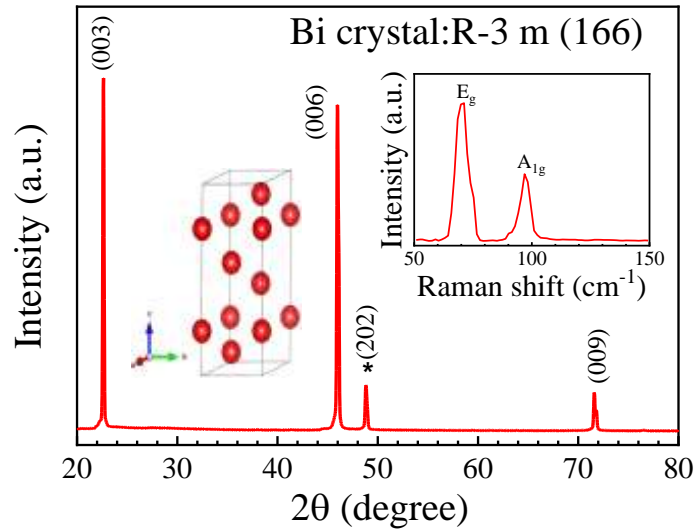

*Fig. 2 Shows the single crystal XRD peaks of Bismuth. The inset on the left shows the rhombohedral unit cell and the inset on the right shows observed Raman peaks. Both the main and the insets are reproduced from ref. [1].*

The single crystalline nature of as-grown Bismuth is confirmed through X-ray diffraction (XRD) spectra performed on a thin flake with peaks aligned along (*00l*) diffraction plane as shown in Fig.2. The Bismuth crystallizes into the rhombohedral unit cell with space group R-3m. Apart from (*00l*) plane peaks observed in Single crystal XRD, an extra peak is also observed and identified as (*202*) plane, which might be due to misalignment of the crystal planes. The XRD was performed multiple times on different flakes mechanically cleaved from the as-grown single crystal. The observed Raman peaks are shown in the inset of Fig. 2 with two peaks of low vibrational frequency identified as $E_g$ and $A_{1g}$ mode. The detailed structural and morphological characterization can be found in the previous report [1]. The magneto-transport measurements show the signature of SdH oscillations in the RH curve, which predicts the topological character of Bismuth. In extension to previous work [1], the same single crystal is investigated for field-dependent RT and angle-dependent RH, to probe the dependence of SdH oscillations on the topological properties of Bismuth.

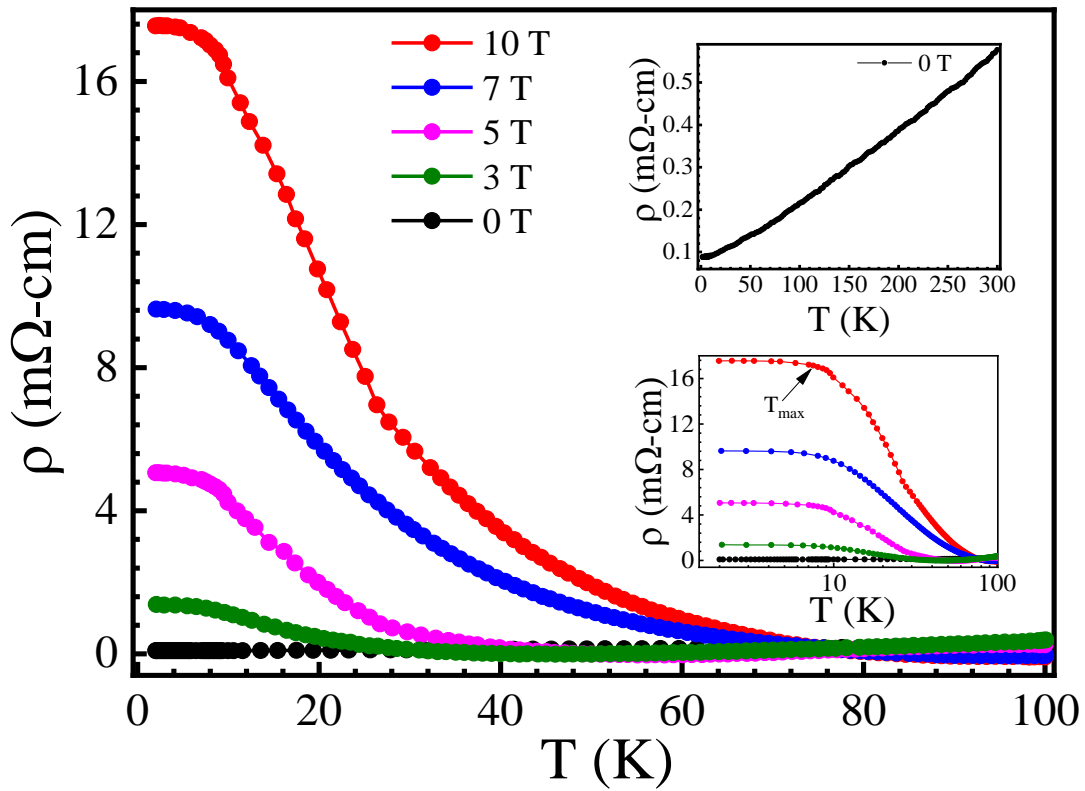

*Fig. 3 Shows the temperature (T) dependence of electrical resistivity (ρ) measured at different transverse magnetic fields. The inset on top shows the metallic behavior of electrical resistivity from 300-2K at zero field and the bottom inset shows resistivity in log scale highlighting the plateau region below 10K.*

Fig. 3 shows the resistivity vs temperature curve of Bismuth in the temperature range from 2 K to 100 K at the magnetic fields 0 T, 3 T, 5 T, 7 T and 10 T. The inset of Fig. 3 shows the RT curve from temperature 300 K down to 2 K at 0 T, as also mentioned in the previous article [1]. At 0 T, the resistivity is found to decrease with the decrease in temperature, which confirms the metallic behavior of Bismuth, as also shown in the previous article [1]. According to Fig. 3, under the applied transverse field the resistance of Bismuth decreases initially with decreasing temperature. However, as the temperature drops further, the resistance begins to rise. This is observed for the RT curve at 3 T, where resistance reduces from 100 K to 42 K and subsequently increases from 42 K down to 2 K. These increases in the RT curve at 3 T are also noticeable for magnetic fields 5, 7, and 10 T. It is important to note that as magnetic field strength increases, the temperature at which resistance curve shows upward trend, shifts towards higher temperature. For 3 T, this temperature is 42 K and for 10 T the same is 92 K. The rise in resistivity at low temperatures indicates an increase in Bismuth bandgap at high magnetic fields. Such an abrupt change in resistivity with increasing field and lowering temperature signifies the deviation from the metallic nature of Bismuth. A similar upturn in resistivity of Bismuth at the different applied transverse fields for Bismuth has been reported [52], which was identified as the field-driven metal-insulator-metal transitions supported by the power law of scaling theory T~(B-B$_c$)$^k$. In contrast to ref. 50, the plateau is observable below T$_{max}$ (see inset of Fig.3) with no hump at T$_{max}$ in our sample.

The increment in resistivity in the low-temperature regime at a higher magnetic field of Bismuth can be explained in terms of its topological surface states [18-20,22]. It is well evidenced in the literature that the topological character of a material gets destroyed in higher magnetic fields. The RT curve at 0 T confirms the metallic character of Bismuth. In contrast, the RT curve at high fields such as 3, 5, 7 and 10 T shows clear deviation from metallic behavior, in particular at lower temperatures. This non-metallic behavior of Bismuth signifies the destruction of topological surface states at high fields. Considering the thermally activated transport as in the intrinsic semiconductor, the thermal activation energy is evaluated by theoretically fitting the measured RT cures at the different magnetic fields by using the relation, $\rho(T) \propto exp(E_g/k_B T)$ where E$_g$ is bandgap and k$_B$ is Boltzmann constant.

Fig. 4 shows the fitting of the natural log of resistivity with respect to 1/T at different magnetic fields, where T is the temperature. It is found that the bandgap of as-grown Bismuth single crystal at 0 T is zero, as its RT curves symbolize the metallic behavior. Further, the theoretical fitting of the RT curve at 3 T predicts a bandgap of 0.018 eV, which further increases to 0.032 eV, 0.075 eV and 0.102 eV respectively at 5, 7 and 10 T. Thus, with an increase in applied magnetic field, the metallic topological character of Bismuth diminishes, and an opening of a gap at topological surface states develops in Bismuth. A similar method has been used to analyse the resistivity upturn in Bi/Pnictide-based crystals [53]. Alternatively, these upturn phenomena in topological materials LaBi [54] and $WTe_2$ [55] are successfully described by a simple two-band model.

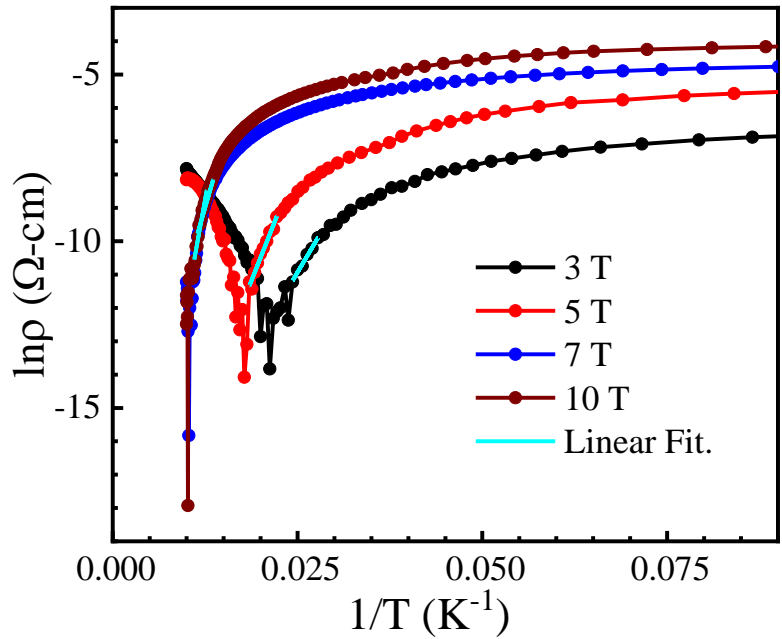

*Fig. 4 Shows the behavior of lnρ as a function of 1/T at different transverse magnetic fields (H).*

Further, the magnetoresistance (MR) of Bismuth is investigated at different angles between the applied magnetic field and the current, which is shown by a schematic in the inset of Fig. 5(a). The detailed investigation of MR at zero degree (Field perpendicular to current direction) is mentioned in the previous article [1], whereas in this article, the angle-dependent MR is investigated at different angles of 30°, 60° and 90°. Fig. 5 (a) shows the angle-dependent MR for angles 60°, 30° and 0° at 2K. The oscillations are evident for 30° and 60° but for 0° the oscillations are damped. It is observed that the MR decreases with a decrease in the angle between the applied field and the current direction. The same is not only minimum but becomes –ve, when the field is parallel to the current direction. Fig. 5(b) and (c) show the angle-dependent MR at 60° and 30° respectively at different temperatures. Fig. 5(b) shows the MR% at 60°, which is nearly around 14500% at 2K and 10T. Further, it is found that the MR% decreases drastically as the temperature is increased to 100 K.

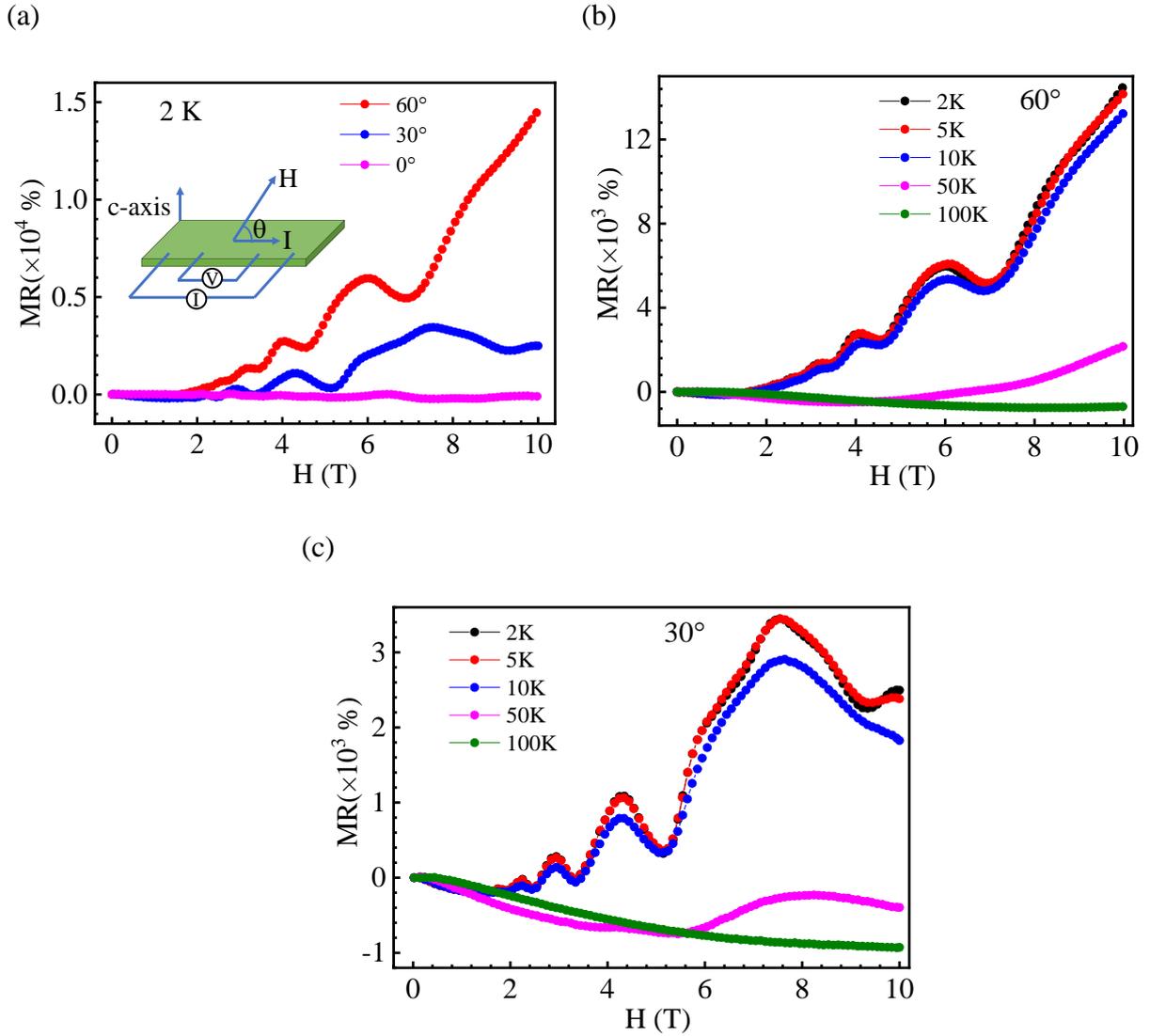

*Fig. 5 (a) Shows the variation of MR% at 2K with respect to the applied magnetic field (H) at different angles. The inset of Fig. 5(a) shows the schematic of field arrangement from normal to sample surface. (b) and (c) Shows the variation of MR% at different temperatures with respect to the applied magnetic field (H) at angles 60° and 30°.*

In Fig. 5 (c), for the applied field at 30°, the MR % is smaller as compared to that of 60°. Also, the quantum oscillation vanishes at 50 K and higher temperatures for both orientations of the field. From Fig. 5(a), it is also observed that, as the inclination angle between the magnetic field and the current is decreased from 60° to 0°, the magnitude of MR also decreases at 2 K.

The signatures of SdH oscillations are observed at higher magnetic fields and low temperatures, which is quite similar to that observed for perpendicular fields in the previous article [1]. Fig 6 (a) and (b) show the SdH oscillations at 60° and 30°, respectively. It is seen from the extracted SdH oscillations, that the harmonicity of these oscillations reduces with decreasing angle i.e., field direction towards the current plane. At 0° i.e. field in a parallel direction with current, the oscillations are damped and even the MR has become slightly –ve. This symbolizes the increment of bulk states contribution over the topological surface states in MR measurements. The SdH oscillations at 60° and 30° are of quantum nature, which is found to be present in various other quantum materials [56,57]. These SdH oscillations are the clear signatures of the topological character of the studied Bismuth single crystal. Interestingly, the same are found to be highly dependent on the angle between the applied magnetic field and the current direction. To extract the exact nature of topology with respect to the angle of the applied magnetic field, the SdH oscillations are theoretically fitted to deduce the Landau level (LL) fan diagram and thereby predict Berry's phase. The corresponding Berry phase at different angles is extracted from SdH oscillations, where the LL index (n) is assigned to the maxima (integer) and minima (Half integer) of the same. The oscillating minima and maxima occur due to crossing of Fermi level by LLs as we increase the magnetic field. The LL index (n) indicate that the Landau levels are shifted due to applied magnetic field so that $n^{th}$ LL is crossing the Fermi level. When the Fermi energy lies in the middle of a LL, there are maximum charge carriers available for conduction resulting minima in resistivity while Fermi level lies in the middle of two LL, minimum charge carriers are available showing maxima in resistivity. This is similar to the methodology being used in previous articles [58,59]. Further, the LL index (n) is plotted with respect to the inverse of the magnetic field and is shown in Fig 7. It represents the LL fan diagram, showing the variation of the LL index at different angles between the applied magnetic field and the current direction. The LL index is found to be dependent on the angle of the magnetic field. All these LL indexes at particular sample orientations are linearly fitted through the Lifshitz-Onsager quantization rule, which is given as $A_F\hbar/eH = 2\pi(n + 1/2 + \beta)$, where $2\pi\beta$ describes the Berry's phase and the theoretical linear fitting is carried out using the origin software [60,61]. The extracted LL index and fitted straight line using the Lifshitz-Onsager quantization rule are found to be in good agreement with each other. The intercept of the fitted straight line describes Berry's phase, and it is found that the values of intercept are 0.503 and 0.882 at 60° and 30°, respectively. Whereas, it is observed that the value of intercept at 90° (field perpendicular to current direction) is 0.531 as is mentioned in the previous article [1]. The obtained values of intercept correspond to $1.006\pi$ and $1.764\pi$ at 60° and 30°, respectively and the same is $1.062\pi$ at 90° [1]. The observed non-trivial $\pi$ Berry phase calculated by the intercept value clearly indicates the topological electrons. The same non-trivial $\pi$ Berry phase was observed in the graphene [62], topological insulators [63], Rashba semiconductor BiTeI [64]. Thus, we confirmed the existence of linear dispersion Dirac fermions in our Bi sample and this suggests the presence of topologically nontrivial charge carriers in Bi. Future quantum oscillation measurements to higher fields are desirable to get a more reliable estimate of the Berry phase. Through the LL fan diagram and corresponding Berry phase calculation, it is clear that the topological character of Bismuth is prominent at 60°, which is slightly suppressed at 90° and diverts at 30°. This study will be useful to find out the topological nature of other materials–by choosing the appropriate applied field angle and current direction.

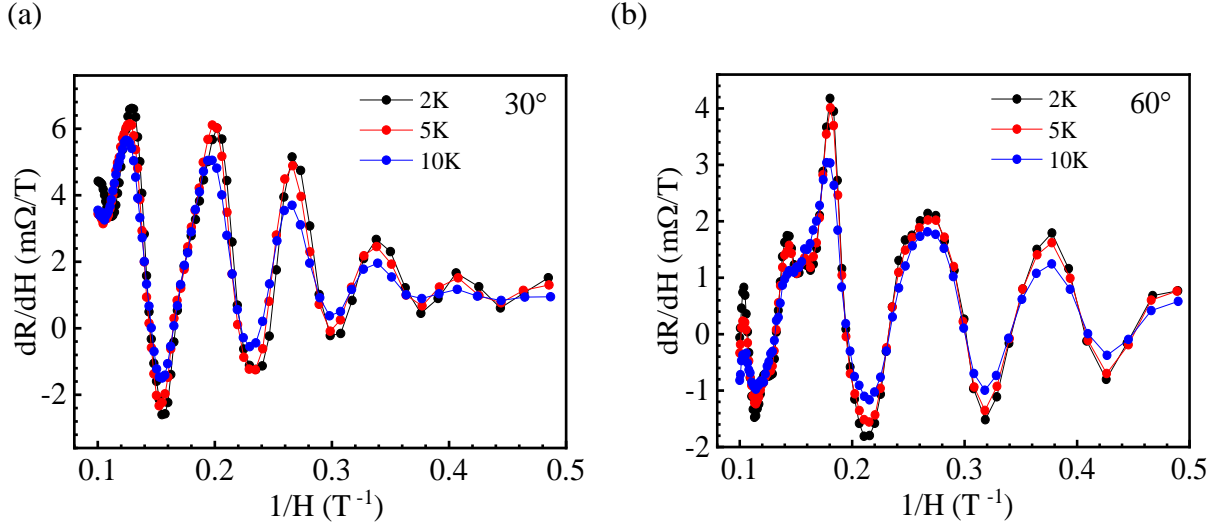

*Fig. 6 Shows the variation of dR/dH as a function of the inverse of the applied magnetic field ($H^{-1}$) at low temperatures. (a) and (b) Represent the SdH oscillation behavior by varying the angle between the applied field and the current at 60° and 30° respectively.*

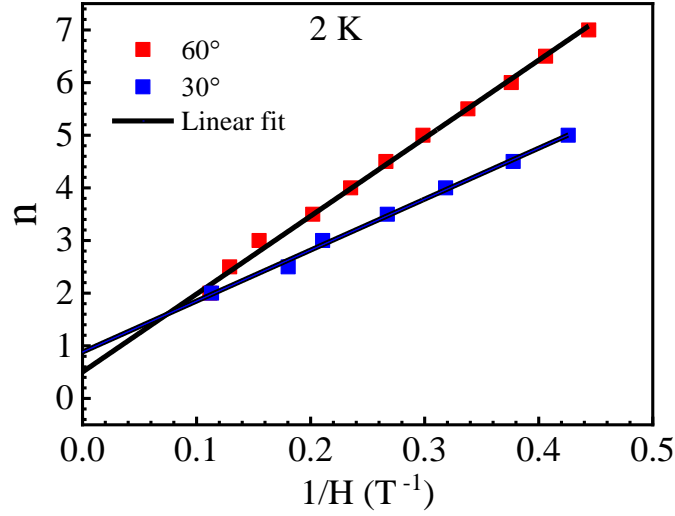

*Fig. 7 Shows the Landau level (n) fan diagram for temperature 2 K, measured for different angles between the inversed applied field (H) and the filled Landau Levels.*

**Conclusion:**

In this article, we investigate the magnetic field and angle-dependent transport properties of Bismuth single crystal, where it is found that there is development of a bulk band gap of 0.102 eV at 10 T, whereas the same is highly metallic in the absence of applied magnetic field. The magnetoresistance is found to be dependent on the angle between the applied magnetic field and the current direction. Both the magnetic field-dependent RT and angle-dependent RH symbolize the presence of surface states in lower and perpendicular magnetic fields. Whereas, there is dominancy of bulk bands over surface states at lower and parallel magnetic fields. The analysis of SdH oscillations indicates that the topological character of Bismuth is prominent at a 60° angle of the field to the current. The theoretically predicted Z2 values also suggested the presence of non-trivial topological surface states.

**Acknowledgement:**

The Director of NPL strongly supports this work and is acknowledged for his interest and encouragement. The authors also thank CSIR for financial support and AcSIR for enrolment as a research scholar in the Ph.D. program. Dr. Geet Awana acknowledges DST India for SERB NPDF Sanction Order No. PDF/2023/003406.

**Data availability statement:**

Data will be made available on reasonable request.

**Declaration:**

The authors declare that they have no known competing financial interests or personal relationships that could have appeared to influence the work reported in this paper.

**Author Contribution:**

N. K. Karn: Draft manuscript preparation, Analysis and interpretation of results

Yogesh Kumar: Draft manuscript preparation, Analysis and interpretation of results

Dr. Geet Awana: Manuscript editing, discussion & literature survey

Dr. V. P. S. Awana: Study conception and design, Supervision, Manuscript Review

All authors reviewed the results and approved the final version of the manuscript.